\documentclass[runningheads,a4paper]{llncs}

\usepackage{amsmath}
\usepackage{amssymb}
\usepackage{bm}
\usepackage{algorithm}
\usepackage{algorithmic}
\usepackage{graphicx}

\usepackage{url}
\urldef{\mailsa}\path|{alfred.hofmann, ursula.barth, ingrid.haas, frank.holzwarth,|
\urldef{\mailsb}\path|anna.kramer, leonie.kunz, christine.reiss, nicole.sator,|
\urldef{\mailsc}\path|erika.siebert-cole, peter.strasser, lncs}@springer.com|

\newcommand{\Exp}{\mbox{\rm E}}
\newcommand{\Prob}{\mbox{\rm Pr}}
\newcommand{\Real}{\mbox{\rm R}}

\newtheorem{assumption}{Assumtion}%[section]

\begin{document}

\mainmatter  % start of an individual contribution

\title{
	Approximation Algorithm \\
	for Cycle-Star Hub Network Design Problems\\
	and Cycle-Metric Labeling Problems
}

\titlerunning{
	Approximation Algorithm 
	for Cycle-Star Hub Network Design Problems
}

\author{Yuko Kuroki$\star$ 
 \and Tomomi Matsui$\star$ 
}
%\author{Yuko Kuroki%
%\thanks{Please note that the LNCS Editorial assumes that all authors have used
%the western naming convention, with given names preceding surnames. This determines
%the structure of the names in the running heads and the author index.}%
%\and Tomomi Matsui}
%%
\authorrunning{Y.~Kuroki and T.~Matsui}

\institute{
\begin{tabular}{cl}
 $\star$  & Department of Industrial Engineering and Economics, \\
& Graduate School of Engineering, \\
& Tokyo Institute of Technology,\\
\end{tabular}
}

\toctitle{Lecture Notes in Computer Science}
\tocauthor{
Yuko Kuroki,
and
Tomomi Matsui
}
\maketitle

\begin{abstract}
We consider a single allocation hub-and-spoke 
	network design problem
	which allocates each non-hub node 
	to exactly one of given hub nodes
	so as to minimize the total transportation cost.
This paper deals with a case 
	in which the hubs are located in a cycle,
	which is called a cycle-star hub network design problem.
The problem is essentially equivalent 
	to a cycle-metric labeling problem.
The problem is useful in the design of networks in telecommunications 
	and airline transportation systems.
We propose a $2(1-1/h)$-approximation algorithm where $h$ denotes 
	the number of hub nodes. 
Our algorithm solves a linear relaxation problem
	and employs a dependent rounding procedure.
We analyze our algorithm by approximating 
	a given cycle-metric matrix
	by a convex combination of Monge matrices.
\end{abstract}

%-----------------------------

\section{Introduction}

In this paper, 
	we propose a $2(1-1/h)$-approximation algorithm 
	for cycle-star hub network design problems with $h$ hubs
	and/or a cycle-metric labeling problem with $h$ labels.

Hub-and-spoke networks arise in 
	airline transportation systems,
	delivery systems and telecommunication systems.
Hub networks have an important role 
	when there are many origins and destinations.
Hub facilities work as switching points for flows. 
In order to reduce transportation costs and set-up costs 
	in a large network,
	each non-hub node is allocated to exactly one of the hubs 
	instead of assigning every origin-destination pair directly.

Hub location problems (HLPs) consist of locating hubs 
	and of designing hub networks so as to minimize the total transportation cost.
%With the single allocation assumptions,
HLPs are formulated as quadratic integer programming problem
	by O'Kelly~\cite{OKELLY1987}, first.
Since O'Kelly proposed HLPs,
	many researches on HLPs %with both single and multiple assignment 
	have been done in various applications
	(see~\cite{ALUMUR2008,CAMPBELL1994,%
CAMPBELL2012,CONTRERAS2015,%
KLINCEWICZ1998,OKELLY1994} for example).

%	e.g., $p$-hub median problems,
%	uncapacitated hub location problems,
%	$p$-hub center problems and hub covering problems. 
%Reviews on HLPs are in O'Kelly and Miller~\cite{OKELLY1994}, 
%	Campbell~\cite{CAMPBELL1994},
%	Klinceswicz~\cite{KLINCEWICZ1998}, 
%	Alumur and Kara~\cite{ALUMUR2008}, 
%	Campbell and O'Kelly~\cite{CAMPBELL2012}.
%	and Contreras~\cite{CONTRERAS2015}.

% Recently, problems which deal with a large scale instance
%	 such as $1000$ nodes are studied.
% Thus, the necessity of researches 
% 	on polynomial time approximation algorithms
% 	is higher than before.
In this paper, we discuss the situation in which  
	the locations of the hubs are given,
	and deal with a problem,
	called a 
	{\em single allocation hub-and-spoke network design problem},
	which finds a connection of the non-hubs to the given hubs
	minimizing total transportation cost.
Sohn and Park~\cite{{SOHN1997},{SOHN2000}} 
	proposed a polynomial time exact algorithm 
	for a problem with 2 hubs
	and proved NP-completness 
	of the problem even if the number of hubs is equal to 3.
Iwasa et al.\@~\cite{IWASA2009}
	proposed a simple $3$-approximation algorithm 
	and a randomized $2$-approximation algorithm
	under the assumptions of triangle inequality.
They also proposed a $(5/4)$-approximation algorithm 
	for the special case where the number of hubs is 3.
Ando and Matsui~\cite{ANDO2011} deal with the case in which
	all the nodes are embedded in a 2-dimensional plane 
	and the transportation cost of an edge per unit flow is
	proportional to the Euclidean distance 
	between the end nodes of the edge.
They proposed a randomized $(1+2/\pi)$-approximation algorithm.
Saito et al.\@~\cite{SAITO2009} discussed 
	some facets of  polytopes corresponding to
	the convex hull of feasible solutions of the problem.

Fundamental HLPs assume a full interconnection between hubs.
Recently, several researches consider 
	incomplete hub networks which arise especially 
	in telecommunication systems
	(see~\cite{CAMPBELL2005a,CAMPBELL2005b,ALUMUR2009,CALIK2009}
	for example). 
These models are useful when set-up costs of hub links 
	are considerably large
	or full interconnection is not required.
%Campbell et al.\@~\cite{{CAMPBELL2005a},{CAMPBELL2005b}} 
%	first relaxed the assumptions of full interconnection 
%	between hubs and introduced the hub arc location problem.
%/*Alumur et al.\@~\cite{ALUMUR2009}
%studied the design of single allocation variants of the existing HLPs ,under the incomplete hub network design.*/
%Calik et al.\@~\cite{CALIK2009} studied a single allocation 
%	hub covering problem over incomplete hub networks 
%	and presented a heuristic algorithm based on a tabu-search technique.
There are researches 
	which assume that hub networks constitute 
	a particular structure such as 
	a tree~\cite{KIM1992,CONTRERAS2009,CONTRERAS2010,DESA2013}, 
	a star~\cite{LABBE2008,YAMAN2008,YAMAN2012}, 
	a path~\cite{MARTINS2015}
	 or a cycle~\cite{CONTRERAS2016}.
%/*Kim and Tcha~\cite{KIM1992}, 
%	Contreras et al.\@~\cite{{CONTRERAS2009},{CONTRERAS2010}}, 
%	and Martin de S\'{a} et al.\@~\cite{DESA2013}
%	studied a design of tree-star network 
%	where the upper-level backbone network 
%	is of tree type and the lower-level local access networks 
%	are of star-type.*/
%Labb\'e and Yaman~\cite{LABBE2008}, 
%	Yaman~\cite{YAMAN2008},
%	Yaman and Elloumi~\cite{YAMAN2012} 
%	studied the design of star-star networks
%	where hubs are directly connected to a central node.
%Martin de S\'{a} et al.\@~\cite{MARTINS2015} consider 
%	the hub line location problem 
%	in which  hubs are connected by means of a line 
%	and the aim is to minimize total service time.
%Contreras et al.\@ \cite{CONTRERAS2016} introduced 
% 	the cycle hub location problem in which the hubs are located in a cycle 
%	and presented an algorithm 
%%	based on the branch-and-cut method and a GRASP meta-heuristic.
%
%\begin{figure}[hbt]
%\begin{center}
%			\includegraphics[width=5cm]{cycle-star.eps}
%			\caption{Cycle-star hub network with $h=5$ hubs.}
%			\label{FIG:cycle-star}
%\end{center}
%\end{figure}

In this paper, 
	we consider a single allocation hub-and-spoke network design problem
	where the given hubs are located in a cycle.
	%(see Figure~\ref{FIG:cycle-star}). 
We call this problem the {\em  cycle-star hub network design problem}.
When the number of hubs is~3,
	the hub network becomes a 3-cycle and constitutes a complete graph.
Thus, the $4/3$-approximation algorithm 
	for a complete 3-hub network proposed 
	in~\cite{IWASA2009} is valid for this special case.
In this paper we propose a $2(1-1/h)$-approximation algorithm
	when a set of $h$ hubs forms an $h$-cycle. 
To the best of our knowledge,
	our algorithm is the first approximation algorithm
 	for this problem.

A single allocation hub-and-spoke network design problem
	is essentially equivalent to a metric labeling problem
	introduced by Kleinberg and Tardos in~\cite{KLEINBERG2002}, 
	which has connections to Markov random field 
	and classification problems that arise 
	in computer vision and related areas. 
They proposed an $O(\log h \log \log h)$-approximation 
	algorithm where $h$ is the number of labels (hubs). 
Chuzhoy and Naor~\cite{CHUZHOY2004} showed that there is 
	no polynomial time approximation algorithm
	with a constant ratio for the problem unless $\mbox{P}=\mbox{NP}$. 
We deal with a cycle-metric labeling problem
	where a given metric matrix is defined
	by an undirected cycle and non-negative edge length.
Thus, our results give an important class
	of the metric labeling problem,
	which has a polynomial time approximation algorithm 
	with a constant approximation ratio.

\section{Problem Formulation} \label{sec:Problem Formulation}

Let $H=\{1,2,\ldots, h\}$ be a set of hub nodes and 
	$N$ be a set of non-hub nodes where $|H|\geq 3$ and $|N|=n$.
This paper deals with a single assignment hub network design problem
	which assigns each non-hub node to exactly one hub node.
We discuss the case in which the set of hubs forms an undirected cycle, 
	and the corresponding problem is called 
	the {\em cycle-star hub network design problem}.
More precisely, 
	we are given an undirected cycle $\Gamma=(H,T)$ defined by 
	a vertex-set $H$ and an edge-set
	$T=\{\{1,2\}, \{2,3\}, \ldots , \{h-1, h\}, \{h,1\}\}$.
In the rest of this paper, 
	we identify hub $i$ and hub $h+i$ when there is no ambiguity.
For each edge $e=\{i, i+1 \} \in T$, 
	the corresponding length, denoted by $c_e$ or $c_{i\, i+1}$ 
	represents a non-negative cost per unit of flow on the edge.
For each ordered pair $(p ,i) \in N \times H$, 
	$c_{pi}$ also denotes a non-negative cost per unit flow 
	on an undirected edge $\{p,i\}$.
We denote a given non-negative amount of flow from a non-hub $p$ 
	to another non-hub $q$ by $w_{pq} (\geq 0)$.
Throughout this paper, we assume that $w_{pp}=0 \; (\forall p \in N)$.
We discuss the problem for finding an assignment of non-hubs to hubs 
	which minimizes the total transportation cost defined below.

When non-hub nodes  $p$ and $q$ $(p \neq q)$
	are assigned hubs $i$ and $j$, respectively,
	an amount of flow $w_{pq}$ is sent along a path
	$((p,i), \Omega_{ij}, (j,q))$ where 
	$\Omega_{ij}$ denotes a shortest path in $\Gamma=(H,T)$
	between $i$ and $j$.
For each pair of hub nodes $(i ,j) \in H^2$, 
	$c_{ij}$ denotes the length of a shortest path $\Omega_{ij}$.
More precisely, cycle $\Gamma$ contains exactly two paths
	between  $i$ and $j$ and 
	$c_{ij}$ denotes the minimum of the lengths of these two paths.
It is easy to see that $c_{ij}=c_{ji}$.
In the rest of this paper,
	 a matrix $C=(c_{ij})$ defined above 
	is called a {\em cost matrix}
	and/or a {\em cycle-metric matrix}.
The transportation cost corresponding to a flow 
	from $p$ to $q$ is defined by
	$w_{pq}(c_{pi}+c_{ij}+c_{qj})$.

Now we describe our problem formally.
First, we introduce a $0$-$1$ variable $x_{pi}$ 
	for each pair  $\{p,i\} \in N \times H$ as follows:
\[	x_{pi}=	\left\{
				\begin{array}{ll}
					1 & (\mbox{$p \in N$ is assigned to $i \in H$}), \\
					0 & (\mbox{otherwise}). 
				\end{array}
			\right.
\]
\noindent
	We have a constraint $\sum_{i \in H}x_{pi}=1$ for each $p \in N$,
	since each non-hub is connected to exactly one hub.
Then, the cycle-star hub network design problem can be formulated as follows:
\begin{alignat*}{6}
&\mbox{\rm SAP: \quad} && \mbox{\rm min.\quad }
&& 
		\sum_{(p,q) \in N^2} w_{pq}
		\left( 
			  \sum_{i\in H} c_{pi}x_{pi}   
			\right.&&\left.+ \sum_{j\in H} c_{jq}x_{qj} 
%		\right) 
		+
%		+ \sum_{(p,q) \in N^2} w_{pq}
%		\left( 
			\sum_{(i,j)\in H^2} c_{ij}x_{pi}x_{qj} 
		\right)  
	\\
&&& \mbox{\rm s.~t.}		
&& \sum_{i\in H} x_{pi} = 1  && (\forall p \in N), \\ 
&&&&&  x_{pi} \in \{0,1\}		&&(\forall (p,i) \in N \times H).\\
\end{alignat*}
%
%This formulation as a quadratic integer problem 
% is introduced by Sohn and Park.
%They considered the situation of complete hub networks.
The above formulation also appears   
	in~\cite{{SOHN2000},{IWASA2009}}.
In case $h=3$, 3-cycle $\Gamma$ is a complete graph,
	and thus the corresponding problem
	is NP-complete~\cite{SOHN2000}.

Next we describe an integer linear programming problem
	proposed in~\cite{{IWASA2009}}, 
	which is derived from SAP 
	by employing the linearization technique 
	introduced by Adams and Sherali \cite{ADAMS1986}.
We replace $x_{pi}x_{qj}$ with $y_{piqj}$.
We have a new constraint  $\sum_{i \in H}y_{piqj}=x_{qj}$ 
	from the equation $\sum_{i \in H}x_{pi}=1$ 
	by multiplying both sides by $x_{qj}$.
We also obtain a constraint $\sum_{j \in H}y_{piqj}=x_{pi}$ in a similar way.
Then we obtain the following $0$-$1$ integer linear programming problem:
\begin{alignat*}{6}
&\mbox{\rm SAPL: \quad} && \mbox{\rm min.\quad }
&& 
		\sum_{(p,q) \in N^2} w_{pq}
		\left( 
			  \sum_{i\in H} c_{pi}x_{pi}   
			\right.&&\left.+ \sum_{j\in H} c_{jq}x_{qj} 
%		\right) 
		+
%		+ \sum_{(p,q) \in N^2} w_{pq}
%		\left( 
			\sum_{(i,j)\in H^2} c_{ij}y_{piqj} 
		\right)  
	\\
&&& \mbox{\rm s.~t.}		
&& \sum_{i\in H} x_{pi} = 1  && (\forall p \in N), \\ 
&&&&& \sum_{j \in H} y_{piqj} = x_{pi}	&& (\forall (p,q) \in N^2 , \forall j \in H, p<q),\\
&&&&& \sum_{i \in H} y_{piqj} = x_{qj}	&& (\forall (p,q) \in N^2 , \forall i \in H, p<q),\\
&&&&&  x_{pi} \in \{0,1\}		&&(\forall (p,i) \in N \times H), \\
&&&&& y_{piqj} \in \{0,1 \}				&&(\forall i \in H).
\end{alignat*}
% We denote the objective function of SAPL 
% 	by $\widehat{\bm{w}}^{\top}\bm{x}+\widetilde{\bm{w}}^{\top}\bm{y}$ for simplicity.

By substituting  non-negativity constraints of all the variables
	for $0$-$1$ constraints in SAPL, 
	we obtain a linear relaxation problem, denoted by LRP.
We can solve LRP in polynomial time by employing 
	an interior point algorithm.

% We have the following linear relaxation problem by substituting the
% 	constrains $0 \leq x_{pi} \leq 1$ for $0$-$1$ constrains $x_{pi} \in \{0,1\}$.
% \begin{alignat*}{6}
% &\mbox{\rm LRP: \quad} && \mbox{\rm min.\quad }
% && 
% 		\sum_{(p,q) \in N^2} w_{pq}
% 		\left( 
% 			  \sum_{i\in H} c_{pi}x_{pi}   
%			\right.&&\left.+ \sum_{j\in H} c_{jq}x_{qj} 
% 		\right) 
% 		+
% %		+ \sum_{(p,q) \in N^2} w_{pq}
% %		\left( 
% 			\sum_{(i,j)\in H^2} c_{ij}y_{piqj}  
% 		\right)  
% 	\\
% &&& \mbox{\rm s.t.}		
% && \sum_{i\in H} x_{pi} = 1  && (\forall p \in N), \\ 
% &&&&& \sum_{j \in H} y_{piqj} = x_{pi}	&& (\forall (p,q) \in N^2 , \forall j \in H, p<q),\\
% &&&&& \sum_{i \in H} y_{piqj} = x_{qj}	&& (\forall (p,q) \in N^2 , \forall i \in H, p<q),\\
% &&&&& 0 \leq x_{pi}	\leq 1		&&(\forall (p,i) \in N \times H), \\
% &&&&& y_{piqj} \geq 0 				&&(\forall(p,q) \in N^2 , \forall (i,j) \in {H^2}, p<q).
% \end{alignat*} 
%%%%%%%%%%%%%%%%%%%%%%%%%%%%%%%%%%%%%%%%%%%%%%%%%%%%%%%%%%%%%%%%%%%%%
\section{Monge Property and Dependent Rounding Procedure}

% \subsection{Monge Property}

First, we give the definition of a Monge matrix.
A comprehensive research on the Monge property 
	appears in a recent survey~\cite{BURKARD1996}.

\begin{definition}\label{defnitonA}
An $m \times n$ matrix $C'$ is a Monge matrix
	if and only if $C'$ satisfies the so-called Monge property
\[
	c'_{ij}+c'_{i'j'} \leq c'_{ij'}+c'_{i'j}\quad\quad\quad
\mbox{\rm for all}\quad1 \leq i <i' \leq m, 1 \leq j < j' \leq n.
\]
\end{definition}

\noindent
Although the Monge property 
	depends on the orders of the rows and columns,
	in this paper, we say that a matrix is a Monge matrix
	when there exist permutations of rows and columns
	which yield the Monge property. 

For each edge $e \in T$, 
	we define a path $\Gamma^{e}=(H,T\setminus \{e\})$ 
	obtained from cycle $\Gamma$
	by deleting the edge $e$.
Let $C^{e}=(c^{e}_{ij})$ be a cost matrix 
	where
	$c^{e}_{ij}$ denotes the length of
	the unique subpath of $\Gamma^{e}$ 
	connecting $i$ and $j$.

\begin{lemma}\label{monge}
For any edge $e=\{\ell, \ell +1\} \in T$, 
	a Monge matrix is obtained from $C^e$ 
	by permuting rows and columns simultaneously
	in the ordering $(\ell+1, \ell+2, \ldots , h,1, 2,\ldots , \ell)$.
\end{lemma}

\noindent
Proof is omitted 
(see~\ref{appendix:proof_of_lemma_monge} 
	or~\cite{BURKARD1996} for example).
%(see~\cite{BURKARD1996} for example).

\smallskip

Next, we approximate a given cost matrix
	(cycle-metric matrix) $C$ 
	by a convex combination 
	of $h$ Monge matrices $\{C^{e} \mid e \in T\}$. 
Alon et al.\@~\cite{ALON1995} considered 
	approximating a cycle-metric matrix 
	by a probability distribution over \mbox{path-metric} matrices, 
	and showed a simple distribution
	such that the expected length of each edge 
	is no more than twice its original length.
The following theorem improves their result
	especially when the size of the cycle 
	(number of hubs) is small.

\begin{theorem} \label{thm:2(1-1/h)}
Let $C$ be a cost matrix obtained from  
	a cycle $\Gamma=(H,T)$ and non-negative edge lengths 
	$(c_e \mid e \in T)$.
Then, there exists a vector of coefficients 
	$(\theta_e  \mid e \in T)$ satisfying
\[
	\theta_e \geq 0 \; (\forall e \in T), \;\;
	\sum_{e \in T}\theta_e=1, 
	\; \mbox{ and } \;
	C \leq \sum_{e \in T} \theta_e C^e 
	\leq 2 \left( 1-\frac{1}{h} \right) C.
\]
\end{theorem}

\begin{proof}
When there exists an edge $e^{\circ} \in T$ satisfying
	$c_{e^{\circ}} \geq (1/2)\sum_{f \in T}c_f$,
it is easy to see that for every pair $(i,j)\in H^2$,
	there exists a shortest path $\Omega_{ij}$ 
	on cycle $\Gamma=(H,T)$	
	between $i$ and $j$ excluding edge $e^{\circ}$. 
Thus, a given cost matrix $C$ is equivalent 
	to the Monge matrix $C^{e^{\circ}}$.
In this case, the desired result is trivial.

We assume that $2c_e < L=\sum_{f \in T}c_f \; (\forall e \in T)$ and  
 introduce a positive coefficient 
	$\theta_e$ for each $e \in T$  defined by
\[
	\theta_e= \frac{c_{e}}{K} 
		\prod_{f \in T \setminus \{e \}}(L-2c_{f})
\]
	where $K$ is a normalizing constant 
	which yields the equality $\sum_{e \in T} \theta_e =1$.
Let $\Omega_{ij} \subseteq T$ be a set of edges in a shortest path 
	in $\Gamma$ between $i$ and $j$.
The definition of the coefficients $(\theta_e \mid e \in T)$ 
	directly implies that for each pair $(i,j) \in H^2$,
\begin{align*}
\sum_{e \in T}\theta_e c^e_{ij} 
&	=\sum_{e \not \in \Omega_{ij}}	\theta_e c_{ij}
	+\sum_{e \in \Omega_{ij}}			\theta_e (L-c_{ij}) 
	=\sum_{e \in T}	\theta_e c_{ij} 
	+\sum_{e \in \Omega_{ij}}
	\theta_e (L-2c_{ij}) \\
&\leq 
	c_{ij} \sum_{e \in T}	\theta_e  
	+\sum_{e \in \Omega_{ij}}
	\theta_e (L-2c_e)  \\ 
&=	
	c_{ij}
	+\sum_{e \in \Omega_{ij}}
	\left(
		 (L-2c_e)  \frac{c_{e}}{K} 
		\prod_{f \in T \setminus \{e \}}(L-2c_{f})
	\right) \\
&= c_{ij}	
	+\frac{\prod_{f\in T}(L-2c_f)}{K}
	\sum_{e \in \Omega_{ij}}c_e
=  c_{ij}	
	+\frac{\prod_{f\in T}(L-2c_f)}{K}
	c_{ij}.
\end{align*}
From the assumption,
	the last term appearing above is positive.
Then, we have 
\begin{align*}
\frac{K}{\prod_{f\in T}(L-2c_f)}
&= \frac{
		\sum_{e \in T}\left( c_e \prod_{f \in T\setminus \{e\}}(L-2c_f) \right)
	}{
		\prod_{f\in T}(L-2c_f)
	}
= \sum_{e \in T} \frac{c_e}{L-2c_e}.
\end{align*}

\noindent
Now we introduce a function
	$f(z_1,\ldots ,z_h)=\sum_{\ell =1}^h z_{\ell}/(L-2z_{\ell})$
	defined on a domain 
	$\{\bm{z}\in [0,L/2)^h \mid z_1+\cdots +z_h=L\}$.
From the convexity and symmetry of variables  of $f$, 
	the minimum of $f$ is attained at $z_1=z_2=\cdots = z_h =L/h$,
	and $f(L/h, \ldots ,L/h)=1/(1-2/h)$, which gives the following inequality
\[
		\sum_{e \in T}\theta_e c^e_{ij} 
	\leq  c_{ij}	
	+\frac{\prod_{f\in T}(L-2c_f)}{K} c_{ij}
	\leq c_{ij}+\left(1 - \frac{2}{h} \right) c_{ij}
	=2\left( 1- \frac{1}{h} \right) c_{ij}.
\]
Since $C \leq C^e \; (\forall e \in T)$,
	it is obvious that $C \leq \sum_{e \in T}\theta_e C^e$.
\end{proof}

%%%%%%%%%%%%%%%%%%%%%%%%%%%%%%%%%%%%%%%%%%%%%%%
%\subsection{Dependent Rounding Procedure}

Next, we describe a rounding technique
	proposed in~\cite{{IWASA2009}}.
We will describe a connection 
	between the Monge matrix 
	and the rounding technique, later,

\begin{algorithm}\label{rounding}
	\textbf{\tt Dependent Rounding $(\bm{x},\bm{y};{\pi}$)}
\begin{description}
\item[\textbf{Input:}] A feasible solution $(\bm{x}, \bm{y})$ 
	of LRP and a total order $\pi$ of the hubs.
\item[\textbf{Step 1:}] Generate a random variable $U$ 
	which follows a uniform distribution defined on $[0,1)$.
\item[\textbf{Step 2:}] Assign each non-hub node $p \in N$ 
	to a hub $\pi (i)$,
	where $i \in{\{1,2,\ldots h\}}$ is the minimum number that satisfies
 $U < x_{p\pi (1)}+ \cdots +x_{p\pi(i)}$.
\end{description}
\end{algorithm}

\begin{figure}[hb]
\begin{center}
			\includegraphics[width=8cm]{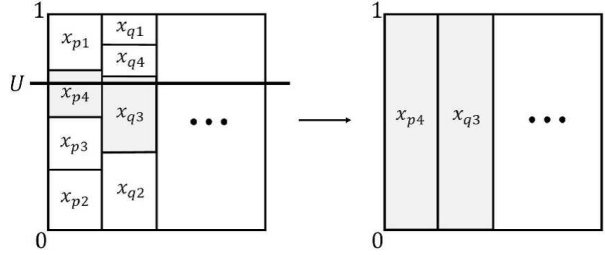}
			\caption{Dependent Rounding $\pi_{\ell}$
		where  $(\pi(1),\pi(2),\pi(3),\pi(4))=(2,3,4,1)$.}
			\label{FIG5.1}
\end{center}
\end{figure}

The above procedure can be explained
	roughly as follows (see Figure~\ref{FIG5.1}).
%Given a feasible solution  $\bm{x}$ of LPR,
For each non-hub $p \in N$,	
	we subdivide a rectangle of height $1$ with horizontal segments
	into smaller rectangles whose heights are equal to 
	the values of the given feasible solution 
	$x_{p \pi(1)}, x_{p\pi (2)},\ldots ,x_{p\pi(h)}$.
Here we note that  $\sum_{i \in H}x_{p \pi(i)}=1 \; (\forall p \in N)$.
We assume that the smaller rectangles are heaped 
	in the order $\pi$.
We generate a horizontal line whose height is equal 
	to the random variable $U$
	and round a variable $x_{pi}$ to $1$ if and only if 
	the corresponding rectangle 
	intersects the horizontal line.

Given a feasible solution $(\bm{x},\bm{y})$ of LRP
  and a total order $\pi$ of $H$, 
  a vector of random variables 
  $X^{\pi}$, indexed by $N \times H$,  
  denotes a solution obtained by 
  {\tt Dependent Rounding}~$(\bm{x},\bm{y}; \pi)$. 
In the following, we discuss the probability
  $\Prob [X^{\pi}_{pi}X^{\pi}_{qj}=1]$.

\begin{lemma}  \label{thm:exp} {\rm \cite{{IWASA2009}}}
Let $(\bm{x},\bm{y})$ be a feasible solution of LRP and
 $\pi$ a total order of $H$.
A vector of random variables $\bm{X}^{\pi}$
  obtained by {\tt Dependent Rounding}~$(\bm{x}, \bm{y}; \pi)$ 
  satisfies that

$
\begin{array}{lll}
\mbox{\rm (1)} & \Exp [X^{\pi}_{pi}]=x_{pi} 
	& (\forall (p,i) \in N \times H), \\
\mbox{\rm (2)} & \Exp [X^{\pi}_{pi}X^{\pi}_{qj}]=y^{\pi}_{piqj} 
 	& (\forall (p,q) \in N^2, \forall (i,j) \in H^2),
\end{array}
$

\noindent
   	where $\bm{y}^{\pi}$ is a unique solution
	of the following system of equalities
\begin{equation}\label{form:NWCsolution}
\sum_{i=1}^{i'} \sum_{j=1}^{j'}  y^{\pi}_{p\pi (i) q \pi(j)}
 = {\rm min}	\left\{ 	
					\sum_{i=1}^{i'} x_{p\pi (i)},
					\sum_{i=1}^{i'} x_{q\pi (j)} 
				\right\}
 	\left( 
			\forall(p,q) \in N^2, \\ \forall(i',j') \in H^2   
	\right).
\end{equation}
\end{lemma}

\noindent
Proof is omitted 
(see~\ref{appendix:DR+NW} or~{\rm \cite{{IWASA2009}}}).
%(see~{\rm \cite{{IWASA2009}}}).

In the rest of this paper, 
	a pair of vectors $(\bm{x}, \bm{y}^{\pi})$
	defined by~(\ref{form:NWCsolution})
	is called a {\em north-west corner rule solution} 
	with respect to $(\bm{x},\bm{y};\pi)$.
When $\bm{x}$ is non-negative and 
	$\sum_{i \in H}x_{pi}=\sum_{i \in H}x_{qi}$ holds,
	the unique solution of~(\ref{form:NWCsolution}) 
	gives the so-called {\em north-west corner rule solution} 
	for a Hitchcock transportation problem
	(Figure~\ref{FIGA.2} shows an example 
	of a north-west corner rule solution 
	where details are given in~\ref{appendix:NWCR}
	or~\cite{IWASA2009}).
Here we note that the above definition is different 
	from the ordinary definition of 
	the north-west corner rule solution,
	which is known as a result of a procedure
	for finding a feasible solution of 
	a Hitchcock transportation problem.
In the rest of this section,
	we describe 
	Hitchcock transportation (sub)problems
	contained in LRP.

\begin{figure}[ht]
%\begin{minipage}{8cm}
%\begin{center}
			\includegraphics[width=8.0cm]{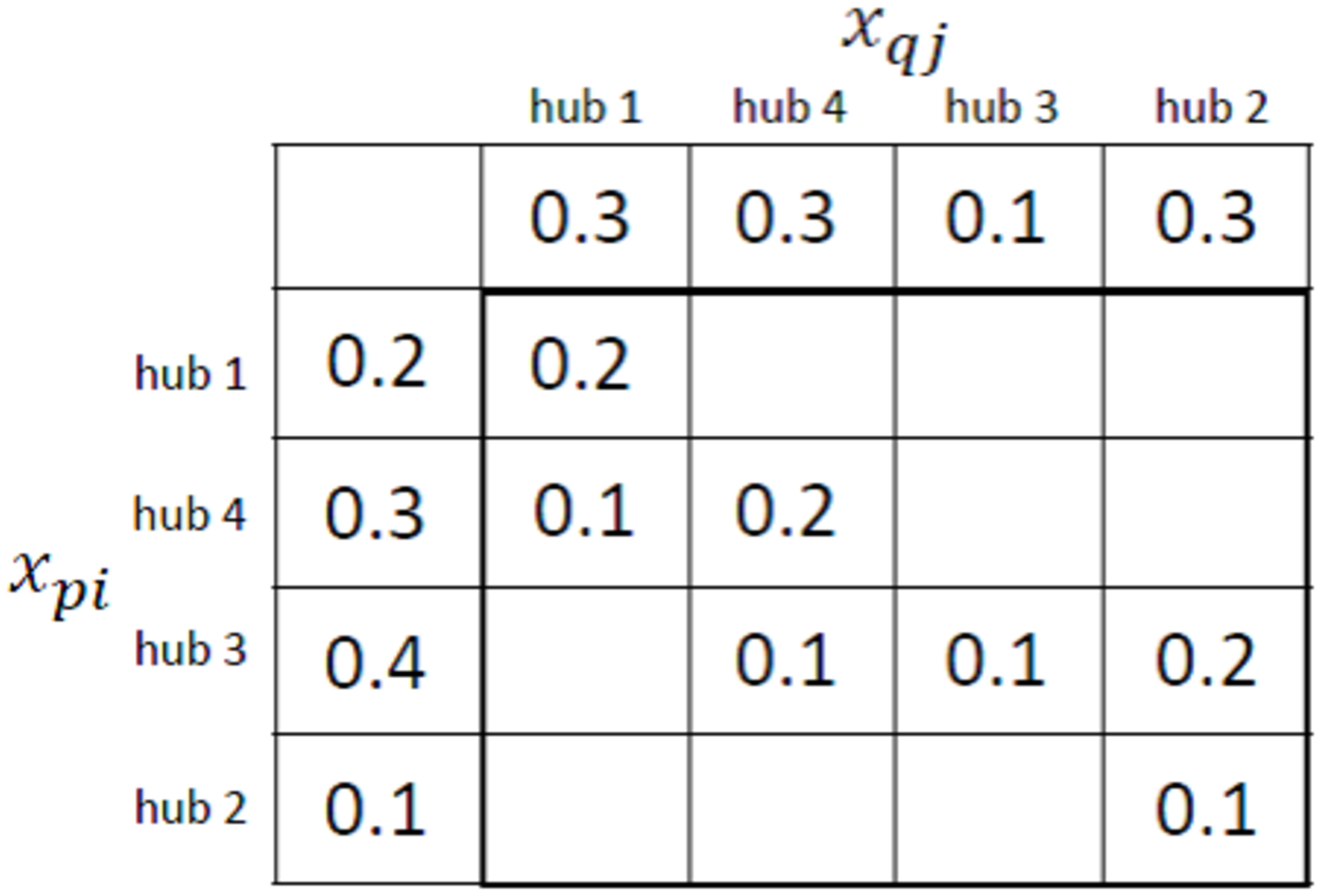}
%\end{center}
%\end{minipage}
%\begin{minipage}{5cm}
%\begin{center}
			\includegraphics[width=4cm]{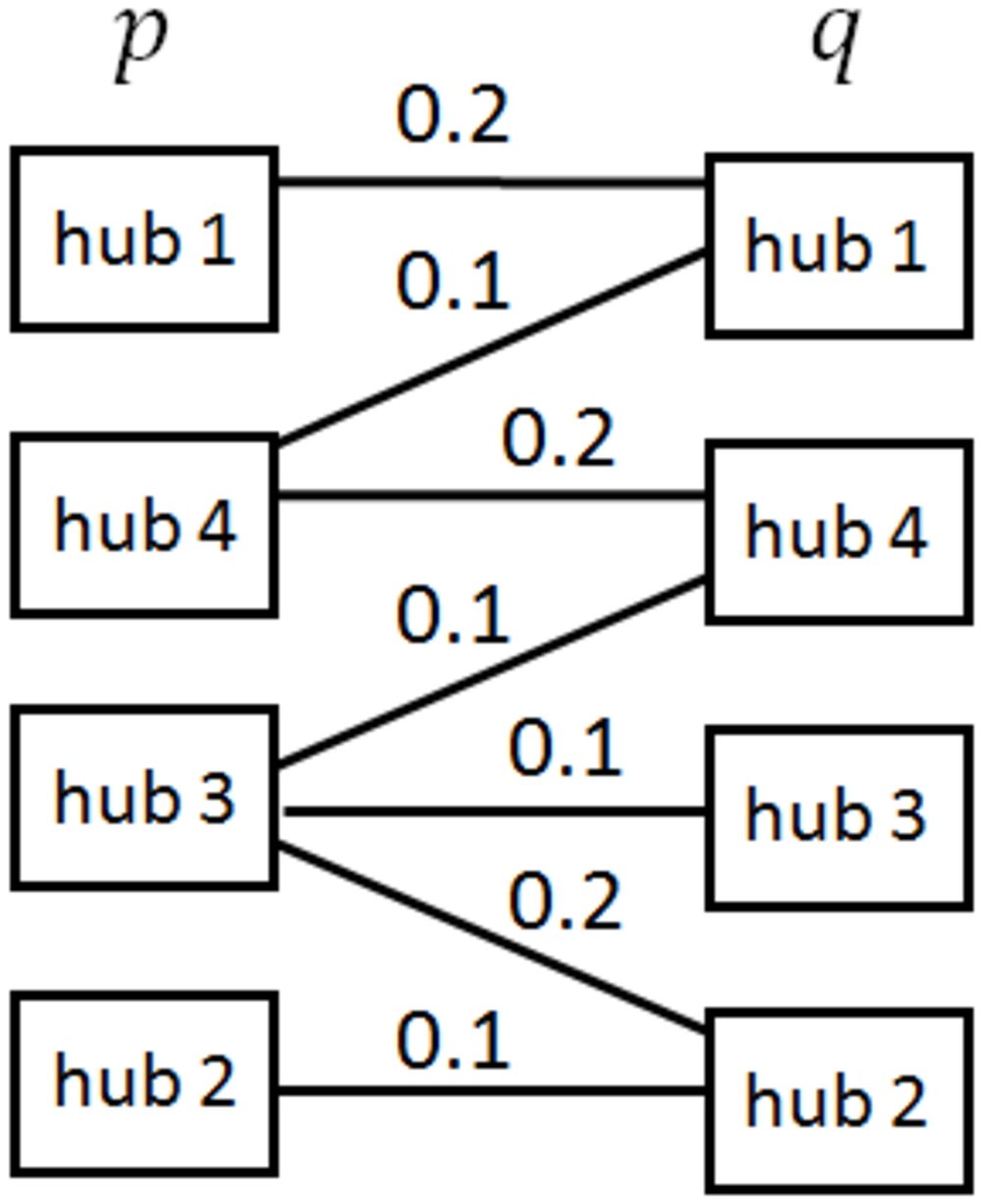}
%\end{center}
%\end{minipage}
			\caption{North-west corner rule solution
			where $(\pi(1),\pi(2),\pi(3),\pi(4))=(2,3,4,1)$.
		In this case, 
		$\Exp [X^{\pi}_{p1}X^{\pi}_{q1}]=0.2$,
		$\Exp [X^{\pi}_{p4}X^{\pi}_{q1}]=0.1$,
		$\Exp [X^{\pi}_{p4}X^{\pi}_{q4}]=0.2$,
		$\Exp [X^{\pi}_{p3}X^{\pi}_{q4}]=0.1$,
		$\Exp [X^{\pi}_{p3}X^{\pi}_{q3}]=0.1$,
		$\Exp [X^{\pi}_{p3}X^{\pi}_{q2}]=0.2$, and
		$\Exp [X^{\pi}_{p2}X^{\pi}_{q2}]=0.1$.
}
			\label{FIGA.2}
\end{figure}
%%%%%%%%%%%%%%%%%%%%%%%%%%%%%%%%%%%%%%%%%%%%%%%%%%%%%%
%\subsection{Hitchcock Transportation Problem}
Let $(\bm{x}^{\circ}$, $\bm{y}^{\circ})$ be a feasible solution
	of LRP.
For any $p \in N$, 
	$\bm{x}^{\circ}_p$ denotes a subvector of $\bm{x}^{\circ}$
	defined by $(x^{\circ}_{p1}, x^{\circ}_{p2}, \ldots , x^{\circ}_{ph})$.  
When we fix variables  $\bm{x}$ in LRP
	to $\bm{x}^{\circ}$,
	we can decompose the obtained problem into $n^2$
	Hitchcock transportation problems
	$\{\mbox{HTP}(\bm{x}^{\circ}_p, \bm{x}^{\circ}_q, C)
		 \mid (p,q) \in N^2\}$
	where 
% $\bm{x}^*_p =(x^*_{p1},x^*_{p2}, \ldots ,x^*_{ph}) \; (\forall p \in N)$
%	and 
 \begin{alignat*}{4}
 \mbox{\rm {HTP}$(\bm{x}^{\circ}_p, \bm{x}^{\circ}_q, C)$: \quad}
& \mbox{\rm min.\quad }
&& 
		\sum_{i \in H} \sum_{j \in H} c_{ij} y_{piqj}  
	\\
& \mbox{\rm s.~t.}		
&&  \sum_{j \in H}  y_{piqj} = x^{\circ}_{pi}  & \quad & (\forall i \in H), \\ 
&&& \sum_{i \in H}  y_{piqj} = x^{\circ}_{qj}  & \quad & (\forall j \in H), \\ 
&&&  y_{piqj} \geq 0			&&(\forall (i,j) \in {H^2}).\\
\end{alignat*} 

Next, we describe a well-known relation 
	between a north-west corner rule solution 
	of a Hitchcock transportation problem and the Monge property.

\begin{theorem}\label{north-west and monge}
If a given cost matrix $C=(c_{ij})$ is a Monge matrix
	with respect to a total order $\pi$ of hubs,  
	then the north-west corner rule solution $\bm{y}^{\pi}$
 	defined by~{\rm (\ref{form:NWCsolution})}
	gives optimal solutions of all the 
	Hitchcock transportation problems
	$\{\mbox{HTP}(\bm{x}^{\circ}_p, \bm{x}^{\circ}_q, C)
		 \mid (p,q) \in N^2\}$.
\end{theorem}

\noindent
Proof is omitted here 
	(see~\cite{{BEIN1995},{BURKARD1996}} for example).
%%%%%%%%%%%%%%%%%%%%%%%%%%%%%%%%%%%%%%%%%%%%%%%%%%%

\section{Approximation Algorithm}

In this section, we propose an algorithm
	and discuss its approximation ratio.
First, we describe our algorithm.

\begin{algorithm}\textbf{\tt  } \label{CycleMetricR}
\begin{description}
\item[Alogorithm~\ref{CycleMetricR}]
\item[Step~1:] Solve the linear relaxation problem LRP
   and obtain an optimal solution  $(\bm{x}^*,\bm{y}^*)$.
\item[Step~2:] 
	For each edge $e \in T$, 
  	execute {\tt Dependent Rounding}~$(\bm{x}^*, \bm{y}^*; \pi^e)$
	where $\pi^e$ denotes a total order 
	$(\pi^e (1),\pi^e (2),\ldots ,\pi^e (h))
		=(\ell+1,\ell+2,\ldots,h,1,2,\ldots,\ell-1,\ell)$.
\item[Step~3:] Output a best solution obtained in Step~2.
\end{description}
\end{algorithm}
In the rest of this section,
	we discuss the approximation ratio.
We define  
%\begin{align*}
%W^*_1= \sum_{(p,q) \in N^2} w_{pq}
%		\left( 
%			  \sum_{i\in H} c_{pi}x^*_{pi}   
%			+ \sum_{j\in H} c_{jq}x^*_{qj} 
%		\right) ,
%		W^*_2=\sum_{(p,q) \in N^2} w_{pq}
%		\left( 
%			\sum_{(i,j)\in H^2} c_{ij}y^*_{piqj}  
%		\right)  
%\end{align*}

\[W^*_1= \sum_{(p,q) \in N^2} w_{pq}
		\left( 
			  \sum_{i\in H} c_{pi}x^*_{pi}   
			+ \sum_{j\in H} c_{jq}x^*_{qj} 
		\right) 
\]
and
\[
W^*_2=\sum_{(p,q) \in N^2} w_{pq}
		\left( 
			\sum_{(i,j)\in H^2} c_{ij}y^*_{piqj}  
		\right)  
\]
\noindent
where $(\bm{x}^*, \bm{y}^*)$ is an optimal solution of LRP.
\begin{theorem} \label{cycle-ratio}
Algorithm~\ref{CycleMetricR} is
	a  $2(1-1/h)$-approximation algorithm.
\end{theorem}
\begin{proof}
Let $z^{**}$ be the optimal value of the original problem SAP
	and $(\theta_e \mid e \in T)$ be a vector of coefficients 
	defined in Theorem~\ref{thm:2(1-1/h)}.
For each $e \in T$, 
%	$(\bm{x}^e,\bm{y}^e)$ denotes an optimal solution of LRP($C^e$).
$(X^{\pi(e)})$ denotes a solution 
obtained by 
	{\tt Dependent Rounding}$(\bm{x}^*,\bm{y}^*;\pi^e)$
and $\bm{y}^{\pi (e)}$ be the north-west corner rule solution 
 	defined by~{\rm (\ref{form:NWCsolution})}
	(where $\pi$ is set to $\pi^e$).
Then we have that
\begin{align*}
&2 \left( 1-\frac{1}{h} \right) z^{**} 
 \geq  2 \left( 1-\frac{1}{h} \right) \mbox{(optimal value of LRP)}
 =     2 \left( 1-\frac{1}{h} \right)(W^*_1+W^*_2) \\
&\geq W^*_1+ 2 \left( 1-\frac{1}{h} \right)W^*_2  
=	W^*_1
		+ 
		\sum_{(p,q) \in N^2} w_{pq}
		\left( 
			\sum_{(i,j)\in H^2}  2 \left( 1-\frac{1}{h} \right) c_{ij}y^*_{piqj}  
		\right)  
	\\
 &\geq \sum_{e \in T}  \theta_e W^*_1
		+ 
		\sum_{(p,q) \in N^2} w_{pq}
		\left( 
			\sum_{(i,j)\in H^2} \sum_{e \in T} \theta_e c^e_{ij}y^*_{piqj}  
		\right)  
	\mbox{\quad (Theorem~\ref{thm:2(1-1/h)}) }
	\\
&= \sum_{e \in T}  \theta_e W^*_1
		+ 
		\sum_{e \in T} \theta_e
		\left( 
			\sum_{(p,q) \in N^2} w_{pq}
			\left( 
				\sum_{(i,j)\in H^2}   c^e_{ij}y^*_{piqj}  
			\right)
		\right)  
	\\
&\geq \sum_{e \in T}  \theta_e W^*_1
		+ 
		\sum_{e \in T} \theta_e
		\left( 
			\sum_{(p,q) \in N^2} w_{pq}
			\left( 
				\mbox{optimal value of HTP}(\bm{x}^*_p,\bm{x}^*_q,C^e) 
			\right)
		\right)  
	\\
%&=	\sum_{e \in T} 
%	\left( 
%		\theta_e 
%		\sum_{(p,q) \in N^2} w_{pq}
%		\left( 
%			  \sum_{i\in H} c_{pi}x^*_{pi}   
%			+ \sum_{j\in H} c_{jq}x^*_{qj} 
%			+ \sum_{(i,j)\in H^2}  c^e_{ij}y^*_{piqj}  
%		\right)
%	\right)  
%	\\
%&\geq
%	\sum_{e \in T} \theta_e 
%	\left(  
%		\sum_{(p,q) \in N^2} w_{pq}
%		\left( 
%			  \sum_{i\in H} c_{pi} x^*_{pi} 
%			+ \sum_{j\in H} c_{jq} x^*_{qj}
%			+ \mbox{optimal value of HTP}(\bm{x}^*_p,\bm{x}^*_q,C^e) 
%		\right) 
%	\right)\\
&= \sum_{e \in T}  \theta_e W^*_1
		+ 
		\sum_{e \in T} \theta_e
		\left( 
			\sum_{(p,q) \in N^2} w_{pq}
			\left( 
				\sum_{(i,j)\in H^2}  c_{ij}^e y^{\pi(e)}_{piqj} 
			\right)
		\right)  
	\mbox{\quad (Theorem~\ref{north-west and monge}) } \\
%&\geq \sum_{e \in T}  \theta_e W^*_1
%		+ 
%		\sum_{e \in T} \theta_e
%		\left( 
%			\sum_{(p,q) \in N^2} w_{pq}
%			\left( 
%				\sum_{(i,j)\in H^2}  c_{ij} y^{\pi(e)}_{piqj} 
%			\right)
%		\right)  \\
&=
	\sum_{e \in T} \theta_e 
	\left(  
		\sum_{(p,q) \in N^2} w_{pq}
		\left( 
			  \sum_{i\in H} c_{pi} x^*_{pi} 
			+ \sum_{j\in H} c_{jq} x^*_{qj}
			+ \sum_{(i,j)\in H^2}  c_{ij} y^{\pi(e)}_{piqj} 
		\right) 
	\right)\\
&=
	\sum_{e \in T} \theta_e 
	\left(  
		\sum_{(p,q) \in N^2} w_{pq}
		\begin{pmatrix}
			  \sum_{i\in H} c_{pi} \Exp [X^{\pi(e)}_{pi}] 
			+ \sum_{j\in H} c_{jq} \Exp [X^{\pi(e)}_{qj}]  \\
			+ \sum_{(i,j)\in H^2}  c_{ij} \Exp [X^{\pi(e)}_{pi}X^{\pi(e)}_{qj}] 
		\end{pmatrix}
	\right)\\
&=
	\Exp \left[
	\sum_{e \in T} \theta_e 
	\left(  
		\sum_{(p,q) \in N^2} w_{pq}
		\begin{pmatrix}
			  \sum_{i\in H} c_{pi} X^{\pi(e)}_{pi}
			+ \sum_{j\in H} c_{jq} X^{\pi(e)}_{qj} \\
			+ \sum_{(i,j)\in H^2}  c_{ij} X^{\pi(e)}_{pi}X^{\pi(e)}_{qj}
		\end{pmatrix}
	\right)
	\right]\\
&\geq
	\Exp \left[
	\min_{e \in T} 
	\left\{  
		\sum_{(p,q) \in N^2} w_{pq}
		\begin{pmatrix}
			  \sum_{i\in H} c_{pi} X^{\pi(e)}_{pi} 
			+ \sum_{j\in H} c_{jq} X^{\pi(e)}_{qj} \\
			+ \sum_{(i,j)\in H^2}  c_{ij} X^{\pi(e)}_{pi}X^{\pi(e)}_{qj} 
		\end{pmatrix}
	\right\}
	\right] \\
&=
	 \Exp [Z] 
\end{align*}

\noindent
where $Z$ denotes the objective value of a solution
	obtained by Algorithm~\ref{CycleMetricR}.
The last inequality in the above transformation is obtained 
	from the equality $\sum_{e \in T} \theta_e=1$ and 
	the non-negativity of coefficients $(\theta_e \mid e \in T)$.
\end{proof}

%%%%%%%%%%%%%%%%%%%%%%%%%%%%%%%%%%%%%%%%%%%%%%%%%
\section{Discussions}

In this paper, 
	we proposed a polynomial time $2(1-1/h)$-approximation algorithm 
	for a cycle-star hub network design problem with $h$ hubs.
Our algorithm solves a linear relaxation problem
	and employs a dependent rounding procedure.
The attained approximation ratio is based on 
	an approximation of a cycle-metric matrix 
	by a convex combination of Monge matrices.

Lastly we discuss a simple independent rounding technique
	which independently connects each non-hub node 
	$p \in N$ to a hub node $i \in H$ with probability 
	$x^*_{pi}$, where $(\bm{x}^*, \bm{y}^*)$ is 
	an optimal solution of LRP.
Iwasa et al.\@~\cite{IWASA2009} showed that 
	the independent rounding technique gives 
	a 2-approximation algorithm
	for a single allocation hub-and-spoke network design problem
	under the following assumption.
\begin{assumption} \label{assume}
A given symmetric non-negative cost matrix $C$ satisfies 
	$c_{ij} \leq c_{ik}+c_{kj} $ $(\forall (i,j,k) \in H^3)$ 
	 and 
	$c_{ij} \leq c_{pi} + c_{pj}$  $(\forall (i,j,p)\in H^2\times N)$.
\end{assumption} 

We also have the following result.

\begin{lemma}
%Under Assumption~\ref{assume},
Under an assumption that 
	$c_{ij} \leq c_{pi} + c_{pj}$  $(\forall (i,j,p)\in H^2\times N)$, 
	a $\left(\frac{3}{2}- \frac{1}{2(h-1)} \right)$-approximation algorithm 
	(for a cycle-star hub network design problem) 
	is obtained by choosing
	the better of the two solutions given by 
	Algorithm~\ref{CycleMetricR} and
	the independent rounding technique.
\end{lemma}

\begin{proof}
Let a random variable $Z_2$ be an objective function value
	with respect to a solution obtained 
	by the independent rounding technique.
Iwasa et al.\@~\cite{IWASA2009} showed that 
	$\Exp [Z_2]\leq 2W^*_1 + W^*_2$.
In the proof of Theorem~\ref{cycle-ratio},
	we have shown that
	$\Exp [Z] \leq W^*_1 + 2(1-1/h)W^*_2$.
By combining these results, we obtain that
\begin{align*}
&\Exp [\min \{Z_2, Z\}]
\leq \Exp \left[\frac{h-2}{2(h-1)}Z_2 +  \frac{h}{2(h-1)}\Exp [Z] \right] \\
&=  \frac{h-2}{2(h-1)} \Exp [Z_2] +  \frac{h}{2(h-1)} \Exp [Z] \\
&\leq 	 \frac{h-2}{2(h-1)}(2W^*_1 + W^*_2)
		+\frac{h}{2(h-1)}\left( W^*_1 + 2\left(1-\frac{1}{h}\right)W^*_2\right)
	\\
&=\left(\frac{3}{2}- \frac{1}{2(h-1)} \right)(W^*_1+W^*_2)
 \leq \left(\frac{3}{2}- \frac{1}{2(h-1)} \right)(\mbox{\rm opt.\@ val.\@ of SAP}).
\end{align*}
\end{proof}

%\section*{Acknowledgement}

%%
%% Bibliography
%%

%% Either use bibtex (recommended), 

%\bibliography{lipics-v2016-sample-article}

%% .. or use the thebibliography environment explicitely

\newpage

%%%%%%%%%%%%%%%%%%%%%%%%%%%%%%%%%%%%%%%%%%%%%%%%%%%%%%%
\renewcommand{\thesection}{Appendix \Alph{section}}
\setcounter{section}{0}

\section{Proof of Lemma~\ref{monge}}\label{appendix:proof_of_lemma_monge}

Here we briefly describe a proof of Lemma~\ref{monge}.

%\begin{lemma}
%For any edge $e=\{\ell, \ell +1\} \in T$, 
%	a Monge matrix is obtained from $C^e$ 
%	by permuting rows and columns simultaneously
%	in the ordering $(\ell+1, \ell+2, \ldots , h,1, 2,\ldots , \ell)$.
%\end{lemma}
%
%Proof is omitted 
%(see the appendix section of~\cite{BURKARD1996} for example).

\begin{proof}
Let $C'$ be a matrix obtained  from $C^e$ 
	by permuting rows and columns simultaneously
	in the ordering $(\ell+1, \ell+2, \ldots , h, 1, 2,\ldots , \ell)$.
We consider each quadruple $(i,i',j,j')$ of indices 
	satisfying $ 1\leq i <i' \leq m, 1 \leq j < j' \leq n$.
Since $C^e$ is symmetric, 
	it suffices to show the following three cases:
\begin{quote}
\begin{description}
\item[\rm (i)] if $i \leq i' \leq j \leq j'$, 
	then 
	$c'_{ij}+c'_{i'j'} = c'_{ii'}+2c'_{i'j}+c'_{jj'}=(c'_{ij'}+c'_{i'j})$,
\item[\rm (ii)] if $i\leq j\leq i'\leq j'$,
	then
	$c'_{ij}+c'_{i'j'} \leq c'_{ij}+c'_{i'j'}+c'_{ji'}+c'_{i'j}=(c'_{ij'}+c'_{i'j})$,
\item[\rm (iii)] if $i\leq j \leq j' \leq i'$,
	then 
	$c'_{ij}+c'_{i'j'} \leq c'_{ij}+c'_{j'i'}+2c'_{jj'}=c'_{ij'}+c'_{ji'}
		=(c'_{ij'}+c'_{i'j})$.
\end{description}
\end{quote}
Thus, we have the desired result. \\ 
\end{proof}

%%%%%%%%%%%%%%%%%%%%%%%%%%%%%%%%%%%%%%%%%%%%%%%%%%%%%%%%%%%%
\section{North-West Corner Rule}\label{appendix:NWCR}

A Hitchcock transportation problem is defined 
	on a complete bipartite graph 
	consists of a set of supply points  $A=\{1,2,\ldots , I\}$ 
	and a set of demand points $B=\{1,2,\ldots , J\}$.
Given a pair of non-negative vectors
	 $(\bm{a},\bm{b}) \in \Real^I \times \Real^J$
	satisfying $\sum_{i =1}^I a_i = \sum_{j=1}^J b_j$ 
	and an $I \times J$ cost matrix $C'=(c'_{ij})$,
	a Hitchcock transportation problem is formulated as follows: 

\begin{alignat*}{4}
 \mbox{\rm HTP}(\bm{a}, \bm{b}, C'):\quad
 & \mbox{\rm min.\quad }
&& 
		\sum_{i=1}^I \sum_{j=1}^J c'_{ij}y_{ij}  
	\\
& \mbox{\rm s.~t.}		
&&   \sum_{j=1}^J y_{ij} = a_i  & \quad & (i\in \{1,2,\ldots, I\}), \\ 
&&&  \sum_{i=1}^I y_{ij} = b_j  & \quad & (j\in \{1,2,\ldots, J\}), \\
&&& y_{ij} \geq 0		
	&&(\forall (i,j) \in \{1,2,\ldots,I\} \times \{1,2,\ldots,J\}),
\end{alignat*}
where $y_{ij}$ denotes the amount of flow 
	from a supply point $i \in A$
	to a demand point $j \in B$.

\begin{algorithm}\label{algorithm:northwest}
	\textbf{Algorithm NWCR}
\begin{description}
\item[\textbf{Step 1:}]
	Set all the elements of matrix $Y$ to $0$ and 
	set the target element $y_{ij}$ to $y_{11}$ (top-left corner).
\item[\textbf{Step 2:}]
	Allocate a maximum possible amount of transshipment to the target element
	without making the row or column total of the matrix $Y$ exceed 
	the supply or demand respectively.
\item[\textbf{Step 3:}]
If the target element is $y_{IJ}$ (the south-east corner element),
	then stop.
\item[\textbf{Step 4:}]
	Denote the target element by $y_{ij}$.
	If the sum total of $j$th column of $Y$ is equal to $b_j$,
	set the target element to $y_{i j+1}$.
	Else (the sum total of $Y$ of $i$th row is equal to $a_i$),
	set the target element to $y_{i+1 j}$.
	Go to Step 2.
\end{description}
\end{algorithm}

We describe north-west corner rule in
	Algorithm~NWCR, which finds
	 a feasible solution of Hitchcock transportation problem
	HTP($\bm{a},\bm{b},C'$).
It is easy to see that the north-west corner rule solution $Y=(y_{ij})$
  satisfies the equalities that 
\[
   \sum_{i=1}^{i'} \sum_{j=1}^{j'} y_{ij}
   = \min \left\{
   \sum_{i=1}^{i'} \alpha_i\;, \;\; \sum_{j=1}^{j'} \beta_j
          \right\} \;\; 
    ( \forall (i', j') \in \{1,2,\ldots,I\} \times \{1,2,\ldots,J\} ).
\]
 Since the coefficient matrix 
	of the above equality system is nonsingular,
	the north-west corner rule solution is a unique solution 
	of the above equality system.
Thus, the above system of equalities has a unique solution 
	which is feasible to 	HTP($\bm{a},\bm{b},C'$).

%%%%%%%%%%%%%%%%%%%%%%%%%%%%%%%%%%%%%%%%%%%%%%%%%%
\section{Proof of Lemma~\ref{thm:exp}}\label{appendix:DR+NW}

We describe a proof of Lemma~\ref{thm:exp} briefly.
See~{\rm \cite{{IWASA2009}}} for detail.

\begin{proof}
(1)  When we introduce an index $i'$ satisfying $\pi(i')=i$, 
	it is easy to see that 
\[ % \textstyle 
	\Exp [X^{\pi}_{pi}]=\Prob [X^{\pi}_{pi}=1]
	= \Prob 
	\left[ 
		\sum_{j=1}^{i'-1}x_{p \pi(j)} \leq U < \sum_{j=1}^{i'}x_{p \pi(j)}
	\right]
	= x_{p \pi (i')}=x_{pi}.
\]

\noindent
(2) We denote 
	$\Exp[X^{\pi}_{pi}X^{\pi}_{qj}]
		=\Prob [X^{\pi}_{pi}X^{\pi}_{qj}=1] $ 
	by $y'_{piqj}$ for simplicity.
Then, the vector $\bm{y}'$ satisfies that
  for any pairs $(p,q) \in N^2 $ 
  and  $(i',j') \in H^2$,
\begin{align*}
& \sum_{i=1}^{i'} \sum_{j=1}^{j'} y'_{p \pi(i) q\pi(j)} 
 = 	\Prob \left[\left[\sum_{i=1}^{i'} X_{p \pi(i)}=1\right] \wedge 
                 \left[\sum_{j=1}^{j'} X_{q \pi(j)}=1\right] \right] \\
&=	\Prob \left[ \left[ U < \sum_{i=1}^{i'} x_{p \pi(i)} \right] \wedge 
                  \left[ U < \sum_{j=1}^{j'} x_{q \pi(j)} \right] \right]  \\
&= \Prob \left[ U < \min \left\{
   \sum_{i=1}^{i'} x_{p \pi(i)}, \sum_{j=1}^{j'} x_{q \pi(j)} 
                           \right\} \right] \\
& =  \min \left\{
   \sum_{i=1}^{i'} x_{p \pi(i)}, \sum_{j=1}^{j'} x_{q \pi(j)} 
                           \right\}.
\end{align*}
From the above, $(\bm{x},\bm{y}')$ satisfies 
	the system~(\ref{form:NWCsolution}) 
	defined by $(\bm{x}, \bm{y}; \pi)$.
The non-singularity of~(\ref{form:NWCsolution}) implies
   \mbox{$\bm{y}'=\bm{y}^{\pi}$}.
\end{proof}

\end{document}